\documentclass[twocolumn,preprintnumbers,amsmath,amssymb]{revtex4}
\usepackage{graphicx}
\usepackage{dcolumn}
\usepackage{bm}

\begin{document}
\title{Non-magnetic impurities to induce magnetism in $\alpha$-PbO crystal structure}
\author{J. Berashevich and A. Reznik}
\address{Thunder Bay Regional Research Institute, 290 Munro St., Thunder Bay, ON, P7A 7T1, Canada}
\address{Department of Physics, Lakehead University, 955 Oliver Road, Thunder Bay, ON, P7B 5E1}

\begin{abstract}
A new route to $d^0$ magnetism is established with help of the first principles methods.
Non-magnetic elements in groups 13 and 14 of the periodic table 
are found to act as the magnetic centers upon embedding in polycrystalline $\alpha$-PbO structure. 
Thus, the local magnetic moment is generated on the impurity site 
(1.0$\mu_B$ and 2.0$\mu_B$ for elements in group 13 and 14, respectively)
due to $p$-orbitals partially filled with electrons whose on-site spin ordering is governed by the first Hund's rule.
The magnetic interactions between impurities are controlled by occupation of the $p$-orbitals such 
as antiferromagnetic ordering (AFM) occurs between impurities 
of 2.0$\mu_B$ while ferromagnetic (FM) between impurities possessing 1.0$\mu_B$. 
In respect to the strength of the magnetic interactions, the atomic radius of impurity is found 
to be a key to tune the wave function tails of localized electrons: with reduction
of the atomic radius the on-site stability of the spin polarized state grows while losing in the long-rang order interactions.
However, it has been shown that a suppression of the long-rang order interactions 
can be compensated by higher impurity concentration that is allowed by shift of the solubility limit to higher magnitude.
\end{abstract}

\maketitle
The research on magnetic semiconductors has intensified in recent years due to requirements imposed by the rapidly
developing field of spintronics \cite{ando}. Originally, magnetic semiconductors were created by
doping of the conventional semiconductors with magnetic ions whose $d$ or $f$ orbitals are partially filled \cite{mahadevan,zunger}.
The enormous attention given to so-called "diluted magnetic semiconductors" has been rewarded 
with the discovery of a mechanism of the 'intrinsic' magnetism in semiconductors –- 
defect-induced magnetism. Initially, in the semiconductors doped with magnetic ions, 
the intrinsic defects were considered only to mediate the magnetic coupling between
localized spins occupying the partially filled $d$ or $f$ orbitals of ions 
thus contributing to the collective magnetism effect \cite{hu,dietl,gohda}. 
However, a better understanding of the defect properties has revealed that the defects
with their $sp$ localized spins are able to generate the magnetic phenomenon themself \cite{coey,venk,podila,dev,peng}. 
The discovery of defect-induced magnetism dubbed as $d^0$ magnetism, 
i.e. magnetism which occurs not due to partially filled $d$ orbitals,
brought new impetus into field of magnetic semiconductors and more importantly in spintronics. 

In order to pursue the spintronics applications, the collective magnetic ordering is 
required to be established between the magnetic centers provided the spin-polarization energy 
of the localized state is large enough for the local magnetic moment to appear above the room temperature.
The stability of the local magnetic moment is defined by the impurity wave function localization 
that unfortunately results in suppression of its tails thus precluding the collective ordering. 
Therefore, success of $d^0$ magnetism in spintronics is recognized to be 
defined by the proper combination of defect/host \cite{zunger} enabling both components. 
In practice, the weak magnetic interactions between the magnetic centers can be 
compensated by their high concentration \cite{coey,venk}. 
However, raising a defect concentration is not always a straightforward solution \cite{zunger}.
For those defects known to induce magnetism: the vacancy \cite{dev,peng,osorio}
and substitutional defects \cite{khan,ye,shen}, the low limit to the defect concentration
even at the most favorable growth conditions is often applied as defined by their formation energy \cite{osorio}. 
Moreover, defects especially in high concentration are not always mechanically
tolerated by the crystal lattice, not to mention that the defect-induced
lattice perturbation may lead to unwanted changes in the electronic properties \cite{zunger}.
Therefore, an idea of intrinsic magnetism requires some efforts to bring it to a level of practical applications.

Our recent finding of new route to $d^0$ magnetism offers an elegant solution and, therefore,
promises a breakthrough in development of magnetic semiconductor \cite{magnet}.
Instead of crystalline systems, the layered materials are proposed to be applied as the semiconductor host.
In crystalline solids, the vacancies only have been considered to establish 
$d^0$ magnetism \cite{dev,peng,osorio,khan,ye,shen} 
because the formation energy of other defects is too high to reach 
the concentrations required for the magnetic percolation to occur.
In respect to the layered systems, the interstitial defects
become a feasible source of unpaired electrons because they are 
incorporated between layers that significantly lowers their formation energy.
In this work we consider the polycrystalline $\alpha$-PbO 
to be semiconductor host for $d^0$ magnetism.
We found that the Pb interstitial defect in $\alpha$-PbO induces the local magnetic moment of 2.0 $\mu_B$ \cite{magnet}. 
The origin of the local magnetic moment upon bonding of the impurity with the host is unique; 
the Pb interstitial of Pb$_i$:$6s^26p^2$ valence shell utilizes its only Pb$_i$:$6s^2$ electrons to be attached to the host 
(through Pb:$6s^2$ electrons as well)
while leaving two unperturbed $6p^2$ electrons on the defect site. The Hund's rule dictates the spin alignment 
of the $6p^2$ electrons (the triplet ground state) that manifests in on-site magnetic moment of 2.0 $\mu_B$ and provides
the high stability of the spin polarized state defined by the spin-polarization energy $E_{pol}$=0.235 eV \cite{magnet}. 
As a result, the Pb atom gains magnetization upon embedding
as the interstitial defect into the $\alpha$-PbO crystal lattice 
(an appearance of the magnetic moment is verified experimentally \cite{expmy}). In analogy with magnetic ions,
magnetism occurs due to partially filled orbitals, 
but here it is due to the $p$ orbital.

\begin{figure}
\includegraphics[scale=0.28]{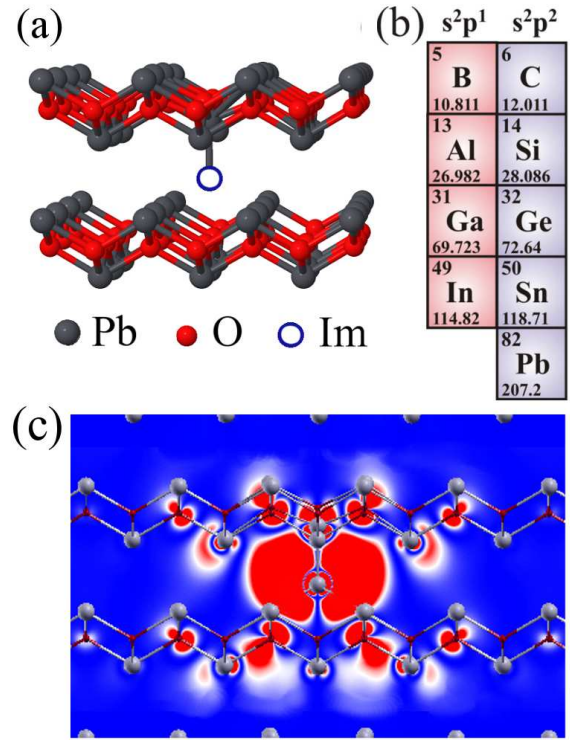}
\caption{\label{fig:fig1} (a) The $\alpha$-PbO crystal structure which contains
the impurity interstitial atom. (b) The list of impurities from group 13 and 14 in periodic 
table used by us
to create the local magnetic moment $\mu$=1.0$\mu_B$ and $\mu$=2.0$\mu_B$, respectively
(the atomic number on the top and the standard atomic weight at the bottom).
(c) The Pb self-interstitial: the spin density map is plotted 
with isovalues of $\pm$0.003 e/\AA$^3$ in Xcrysden for the energy range 
($E_D+E_V$)$\pm$0.15 eV. It demonstrates alignment of electrons at 
the impurity site for which defect tails can be traced up to the last oxygen atom shown.}
\end{figure}

The unique mechanism of bonding which utilizes only $s^2$ valence
electrons allow to extend a choice of host
and the impurity to several candidates.
Because family of the $\alpha$-PbO lattice shows
the lone pair $s^2$ valence shell electrons as a common feature, 
in principal any of those systems can be used as the host. 
In fact, $\alpha$-PbO crystal structure is well recognized in superconductivity \cite{lee}: Fe-pnistides 
(basics are FeSe, FeAs), cuprate (basics are CuO, CuS), and lanthanum compounds (basics are LaF, LaO).
However, among others the $\alpha$-PbO compound is seems to be the best candidate 
due to its wide band gap. For the wide band systems, impurity induces 
the localized defect states inside the band gap which location may vary 
with impurity choice. Any chemical elements possessing partially filled $p$ valence shell 
can gain magnetic properties upon embedding into the $\alpha$-PbO crystal lattice. 
Thus, elements of the same valence 
shell as Pb atom, i.e. belonging to group 14th of the periodic table 
(see Fig.~\ref{fig:fig1}(b)), are expected to induce the local magnetic moment 2.0$\mu_B$.
Following the same principal, the chemical elements of $s^2p^1$ valence shell from group 13
of the periodic table would work as magnetic centers characterized by
the local magnetic moment 1.0$\mu_B$. 
Since we can expect both, a difference in the atomic radius of 
impurity and occupation of the $p$ orbital, to control
the magnetic behavior, focus is on possibility to tune magnetism with different impurities 
looking for a proper combination impurity/host allowing to reach d$^0$ ferromagnetism.
If our hypothesis is proven true, this approach can open a wide perspective
to design the desired magnetic behavior in the $\alpha$-PbO semiconductor
by generating a network of the interstitial defects 
acting as the magnetic centers. The feasibility to generate such network 
is considered through the thermodynamics of the defect formation. 

In our study we applied the generalized gradient approximation (GGA)
with the PBE parametrization \cite{perdew} provided by WIEN2k 
package for the density functional calculations \cite{wien}
(augmented plane wave + local orbitals approach). 
The Pb:$5p,5d,6s,6p$ and O:$2s,2p$ electrons
have been treated as the valence electrons (the energy cutoff was -8 Ry). The supercell approach (RKmax=7)
with sufficiently large supercell of 108-atom size (3$\times$3$\times$3 array of the primitive unit cells)
has been used for single impurity calculation while the 190-atom size (4$\times$4$\times$3)
supercell for the interacting defects. For integration of the Brillouin-zone,
the Monkhorst-Pack scheme of the 5$\times$5$\times$4 (or 4$\times$4$\times$2) k-mesh was applied.
In application to the unpaired electrons, GGA often fails to perform the localization of the defect wave function 
due to an electron self-interaction error \cite{Avezac} that has been examined 
here with the Hartree-Fock (HF) approach applied directly to the unpaired electrons.
Moreover, it is known that when the band gap size is underestimated by GGA,
the hole-carrying impurity orbital may appear above the bottom of the 
conduction band thus inducing
the spurious long-range order interactions \cite{zunger}. 
For the lattice parameters optimized with GGA, the band gap is 1.8 eV 
(very close to the experimental value \cite{exp})
while it found to shrink by 0.22 eV when the experimental lattice parameters are considered \cite{defects}. 
Such gap deviation originates as a result of the interlayer distance
mismatch to occur upon lattice optimization performed with GGA \cite{defects}. 
Since in the $\alpha$-PbO crystal structure the band gap size is controlled 
by the interlayer interactions of the Pb:$6s^2$ electrons, 
application of GGA to optimization of the lattice parameters 
through overestimation of the interlayer distance causes the band gap to increase.
In order to prevent the "spurious" effect, calculations of the electronic property are performed
for the lattice parameters optimized with GGA as it gives the better agreement of the band gap size 
with the experimental data. 
On other hand, to preclude the defect formation energies
to be underestimated, the experimentally determined interlayer distance
has been used for those calculations. The formation energies of the interstitials have been evaluated 
for the vacuum conditions (details on the formation energy simulations are presented in Ref. \cite{def}).

\begin{figure}
\includegraphics[scale=0.70]{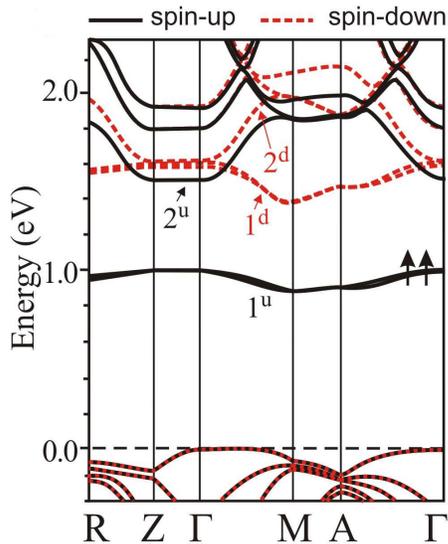}
\caption{\label{fig:fig2} The band diagram for the $\alpha$-PbO crystal structure
containing the Pb self-interstitial defect: $1^u$ and $1^d$ are the bands formed by the $p$-localized electrons 
Pb:$6p^2_{x+y}$, while $2^u$ and $2^d$
are the antibonding orbitals of the Im-Pb bond \cite{magnet}.}
\end{figure}

In respect to origin of the local magnetic moment in the $\alpha$-PbO compound 
on site of the Pb interstitial \cite{magnet}, 
our study had revealed that the Pb interstitial combines the
advantages of the vacancies \cite{coey,venk,podila,dev,peng}
and magnetic ions \cite{mahadevan,zunger}. 
The high spin-polarization energy is observed for Pb interstitial
due to the spin ordering of $6p^2_{x+y}$ to be 
governed by the Hund's rule as for magnetic ions. 
At the same time a hybridization of impurity state with the host lattice
results in the extended defect tails promising to induce
the long-range order interactions. As shown in Fig.~\ref{fig:fig1}(c),
the defect tails appear in upper and lower layers, they are extended up to
seven nearest-neighbors and show the higher spin localization at the oxygen atoms.
Although the impurity interacts with the top layer through bonding while with bottom
layer only through the hybridization interactions, the defect tails
are observed to be more pronounced at the bottom layer. It occurs because the Pb interstitial
is tightly sandwiched between layers that results in its strong hybridization
with the bottom layer. 

\begin{figure*}
\includegraphics[scale=0.70]{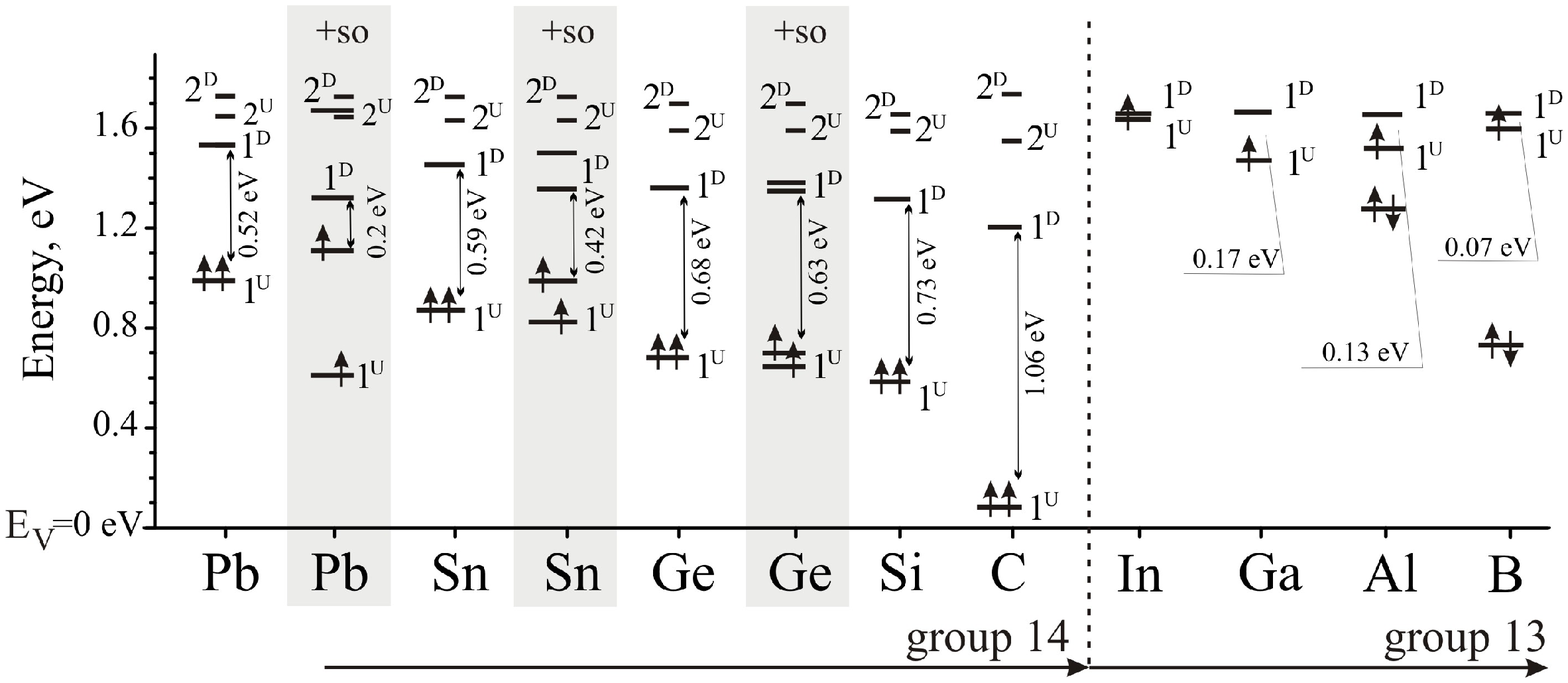}
\caption{\label{fig:fig3} The energetic location of the $1^u$, $1^d$, $2^u$ and $2^d$
bands relative the top of the valence band $E_V$ and splitting of the 
$1^u$ and $1^d$ bands.}
\end{figure*}

The redistribution of the spin density from Pb interstitial site to
the host lattice explains the on-site stability of the triplet state ($E_{pol}$=0.235 eV) 
to be lowered in comparison to the magnetic impurities of $d$ or $f$ types \cite{janak}
known to exhibit the localized nature of the unpaired electrons. 
We expect a hybridization with the host to be a key to tune the magnetic behavior 
when the atomic radius of the impurity is reduced. 
In this work we track an alteration in the electronic properties of 
the host upon replacement of the Pb interstitial with different impurities 
through a behavior of the impurity associated bands 
depicted in Fig.~\ref{fig:fig2} as $1^u$, $1^d$, $2^u$ and $2^d$.
The $1^u$ and $1^d$ bands are those induced by $p$-localized electrons 
(the spin-up band $1^u$ is occupied by $p$ electrons from the impurity valence shell,
while spin-down band $1^d$ is empty), the $2^u$ and $2^d$ bands are 
antibonding orbitals of the impurity-host bond (Im-Pb).
An appearance of all four bands inside the band gap upon changing the impurity 
type is shown in Fig.~\ref{fig:fig3} (absence of $2^u$ and $2^d$ bands 
is referred to the antibonding orbitals outside of the band gap).
Other important parameters such as impurity atomic radius $R_{Im}$, length of the impurity-host bond
Im-Pb, spin-polarization energy $E_{pol}$ 
and splitting of the $1^u$ and $1^d$ bands ($E_1-E_2$)
are disclosed in Table~\ref{tab:table1}.

The common trends on formation of the local magnetic moment as a function 
of the impurity atomic radius are investigated
based on elements from group 14 with
$s^2p^2$ valence shell generating the local magnetic moment 2.0$\mu_B$.
It was found that reduction in atomic radius of impurity leads to shift of both impurity bands $1^u$ and $1^d$
($1^u$ is occupied by two electrons) towards the valence band ($E_V$)
but in the same time the gradual enhancement of the splitting of these bands occurs.
The $2^u$ band is also shifted towards the lower energy due to shortening of the Im-Pb bond, 
while the energetic position of the $2^d$ band deviates in narrow energy range.
The enhanced splitting of the $1^u$ and $1^d$ bands (see ($E_1-E_2$) in Table~\ref{tab:table1})
indicates a gain in stability of the spin-polarized state and, indeed,
a raise in the spin-polarization energy $E_{pol}$ is observed. We found that for the Si and C impurities
which electronic interactions with the bottom layer are reduced because of shortening of 
the Im-Pb bond, 
the on-site stability already approaches the magnitudes known for the magnetic ions \cite{janak}. 

The effect of the spin-orbit coupling (+so) on
the defect bands splitting has been investigated with GGA+so 
for the heavier elements (see Fig.~\ref{fig:fig3}). It is found that for Pb impurity
the spin-orbit coupling breaks degeneracy of the 
$1^u$ and $1^d$ levels resulting in their splitting by 0.52 eV.
Although, such large splitting is responsible for reduction of ($E_1-E_2$), 
the triplet state remains stable that is confirmed by exhibition 
of the local magnetic moment in experiment at room temperature \cite{expmy}.
The spin-orbit effect is less 
pronounced for the Sn impurity for which the spin-orbit splitting is 
reduced to 0.19 eV. For elements with smaller atomic radius, 
the spin-orbit coupling can be neglected: 
it is 0.07 eV for Ge atom and decreases further down for impurities of smaller radius. 

We also have examined an effect of the electron self-interaction error \cite{Avezac}
on magnetic behavior of the Pb impurity through application of the HF approach 
directly to the unpaired electrons.
The spin-polarization energy is found to increase more than twice to
$E_{pol}$=0.490 eV as a result of 
enhancement of the splitting of the $1^u$ and $1^d$ bands to 1.12 eV: 
the $1^u$ orbital is shifted towards the valence band while $1^d$ towards the conduction band
by $\sim$ 0.4 eV each. However, the opposite effect is observed 
when the experimental lattice parameters are taken into account because a reduction in the interlayer distance
results in enhancement of impurity hybridization with the opposite host layer. 
When both are applied, the compensation effect is developed causing a reduction 
of the band splitting to ($E_1-E_2$)=0.68 eV. 
This value is very close to that found with GGA (see Table ~\ref{tab:table1})
that indicates the reliability of GGA for this task. 

The chemical elements from group 13 in the periodic table are also 
found to be able to act as the magnetic impurity 
forming the local magnetic moment 1.0$\mu_B$ induced by
$p$ electron occupying the $s^2p^1$ valence shell. 
Since the $2^u$ and $2^d$ bands appears in the conduction band, 
they are not presented in Fig.~\ref{fig:fig3}. 
In contrast to $s^2p^2$ impurities, 
the $1^u$ orbital occupied by single unpaired electron is found to appear very close 
to the conduction band. This causes the significant 
defect wave function delocalization and, therefore,
much weaker splitting of the $1^u$ and $1^d$ bands.
Thus, the In atom possesses almost zero splitting. The expected growth of ($E_1-E_2$)
is observed for the Ga impurity for which splitting 
reaches 0.17 eV. Because of small splitting, electron from $1^u$ leaks to $1^d$
inducing reduction of the local magnetic moment to 0.96$\mu_B$. 
The spin-orbit coupling effect is weakly pronounced for the Ga impurity 
causing a negligible reduction of ($E_1-E_2$) by 0.015 eV.
Although further increase in the splitting of the $1^u$ and $1^d$ bands
has been expected for the Al impurity, the 
modification of the bonding mechanism has discontinued such trend.
The Al atom is attached to the Pb atom from the bottom layer instead of 
top layer shown in 
Fig.~\ref{fig:fig1} (a). It results in shift of the Im-Pb bonding orbital
from valence band into the band gap. 
The splitting of the $1^u$ and $1^d$ bands 
is found to decrease to 0.13 eV that induces a further reduction of 
the local magnetic moment to 0.78$\mu_B$. For the B atom as the magnetic impurity, 
a suppression of the ($E_1-E_2$) is even stronger leading to disappearance of 
the local magnetic moment. 
Therefore, for new bonding mechanism, the opposite trend is observed:
the spin-polarization energy $E_{pol}$ defining
the stability of the local magnetic moment decreases with reduction of the atomic radius.
The general conclusion, among chemical elements from group 13 in the periodic table
only Ga impurity promises some advantages for d$^0$ magnetism.

\begin{table}
\caption{\label{tab:table1} 
The stability of the spin-polarized state determined within GGA calculation as a function of 
impurity atomic radius $R_{Im}$ \cite{slater}:
the spin-polarization energy $E_{pol}$ and the energy splitting of the $1^u$ and $1^d$ bands
($E_1-E_2$). Im-Pb is a length of the bond to be formed between impurity and the host.}
\begin{tabular}{c|c|c|c|c}
Im & $R_{Im}$, \AA & Im-Pb, \AA & $E_{pol}$, eV & ($E_1-E_2$), eV \\
\hline 
Pb & 1.81 & 2.90 & 0.235 & 0.523 \\
Sn & 1.72 & 2.86 & 0.258 & 0.585 \\
Ge & 1.52 & 2.70 & 0.306 & 0.680 \\
Si & 1.46 & 2.65 & 0.338 & 0.734 \\
C & 0.90 & 2.30 & 0.538 & 1.058 \\
In & 2.00 & 3.06 & 0.000 & 0.000 \\
Ga & 1.81 & 2.92 & 0.051 & 0.173 \\
Al & 1.82 & 2.71 & 0.003 & 0.131 \\
Al & 1.17 & 2.71 & 0.000 & 0.000 \\
\end{tabular}
\end{table}

In respect to formation energy of defect, 
reduction in the atomic radius of impurity leads to 
less distortion within the $\alpha$-PbO crystal structure:
in order to accommodate the Pb interstitial the layers of the host move apart
while for impurity of smaller radius the interlayer distance is preserved.
As a result, we have observed a reduction of the defect formation energy:
1.23 eV is found for Pb interstitial (for details on calculation of the formation 
energy see \cite{def}), 0.79 eV for the Sn impurity,
just above zero for Ge, Si and Ga, and it is becoming negative for C impurity.
The low formation energy promises not only the better mechanical tolerance of the
host lattice to defects but also the higher finite defect concentration to be reached. Thus, thermodynamically granted 
defect concentration for the negative formation energy
can be as high as number of the sites available for bonding $\sim$$10^{22}$ cm$^{-3}$.
The formation energies are found to drop down 
when impurity is placed on surface of single crystal that opens a way to 
perform a doping of the nearest-neighbors sites
(for example by almost 1.0 eV for the Pb interstitial \cite{def}). 
Although the impurity of small atomic radius 
shows a better on-site stability of the local magnetic moment and the low formation energy,
but stronger localization of the defect wave function implies shorter defect tails that 
would influence the long-order interactions.

The collective magnetic ordering may only occur when two impurities 
are close to each other to establish the magnetic coupling of their localized spins.
In this respect the long range order interactions play the essential role.
The magnetic coupling between impurities has been simulated for the system
containing two interstitials of $s^2p^2$ valence shell. The $p^2_{x+y}$ electrons
localized on impurities have been aligned on-site, while their inter-site
ordering has been switched from antiferromagnetic to ferromagnetic in order to
evaluate $E_{M}=E_{AFM}-E_{FM}$.
The $6p^2_{x+y}$ state is exactly half filled and, therefore, if 
the localized electrons of two interacting impurities 
are ferromagnetically coupled (the total magnetic moments is 4 $\mu_B$), 
the inter-site virtual hopping is not allowed \cite{zunger,dev}.
The virtual hopping is supported only for AFM coupling 
(the total magnetic moment for two interacting impurities is zero)
and because it lowers the total energy, AFM becomes the ground state.
For two Pb interstitials placed on distance 4.0 \AA\
we found that $E_{M(Pb-Pb)}$=-0.96 eV (a negative sign indicates the AFM ground state)
while for two interacting C interstitials it is reduced to $E_{M(C-C)}$=-0.38 eV.
The electronic interactions and the magnetic coupling between defects exponentially decrease
with defect separation. Thus, for two defects placed on 
a distance 12.5 \AA\ (for this calculation the size of the supercell was 4$\times$4$\times$3), 
the energy difference between AFM and FM states is drastically supressed 
to $E_{M(Pb-Pb)}$=-0.0056 eV and
$E_{M(C-C)}$=-0.0023 eV for the Pb and C interstitials, respectively. 
These data prove that the larger is the atomic radius of impurity, the stronger is the 
hybridization of the impurity state with the host lattice being responsible 
for extension of the defect tails.
To establish the magnetic percolation, 
a reduction in the inter-impurity coupling to occur for the impurities of smaller atomic radius
can be compensated by the impurity concentration which can be increased 
due to a shift of the
thermodynamic limit of defect formation to the higher magnitudes.
For example, since the C interstitial possesses the negative formation energy, potentially it can be induced 
on the nearest-neighbouring sites. 
For this case, the theoretical limit of the exchange interaction strength defined by $E_{M(C-C)}$=-0.38 eV
can be achieved.

Although the impurities of $s^2p^2$ valence shell show the interesting physics,
but their inter-site AFM ordering and appearance of the defect states deep inside 
the band gap (the coupling of the impurity state to the band like state is essential 
to support the spin-polarized carrier transport) make them inappropriate for spintronics applications.
The impurities from group 13 have a better fit to the requirements imposed by spintronics:
the $1^u$ and $1^d$ bands both couples to the conduction band 
and the FM ground state should be granted for interacting impurities.
Since the Ga impurity has shown the highest potential 
due to its high spin-polarization energy, here we focus on development of 
the magnetic interactions between two Ga impurities.
We found that for two defects placed on 
a distance 4.0 \AA, the strength of the magnetic coupling is defined by
$E_{M(Ga-Ga)}$=1.08 eV (a positive sign indicates the FM ground state).
The distance between impurities has been increased twice 
that corresponds to the realistic defect concentration $x$=2.5$\%$. 
To simulate the worse-case scenario,
the impurities have been attached to the 
opposite layers. In this case, the overlap of the defects tails 
is weakest as the defect tails for small atomic radius impurity 
are stronger pronounced in the layer the impurity is attached to. 
Secondly, instead of straight line location, the impurities have been placed obliquely 
that also reduces the interaction of their defect tails 
(oblique line involves the lead atoms while the defect tails are stronger 
on the oxygen atoms to be on the straight line shown in Fig.~\ref{fig:fig1}(c)).
For such impurity location, the exchange interaction strength are accounted by 
$E_{M(Ga-Ga)}$=0.04 eV (a distance between impurities was 8.66 \AA).
In order to roughly estimate the Curie temperature $T_C$, the simplified mean-field approximation for the Heisenberg model 
can be applied as $T_C=2/3k_BE_{M}$ \cite{curie} ($k_B$ is the Boltzmann constant).
The Curie temperature is found to be just above 300 K for 2.5$\%$ of impurity
but this value is rather underestimated 
as the worse-case scenario on impurity location has been applied.

In summary, we propose to generate the local magnetic moment in compounds 
of $\alpha$-PbO crystal structure by its doping with non-magnetic
impurities belonging to group 13th or 14th of the periodic table.
In analogy with magnetic ions, the magnetic moment origin is due to
partially filled orbitals, but instead of $d$ or $f$ types it is being of $p$-type. The 
partial occupation of $p$ orbital appears as a result of the unique bonding of
impurities with the host lattice: the original partial occupation of $p$ orbital of impurity 
is preserved after its bonding to the host. 
We found that the magnetic behavior of the dopants depends on their atomic radius:
dopants with the smaller radius are found to establish the higher on-site stability of the localized spins.
For the Si and C impurities, the on-site stability reaches such a high magnitude that
it becomes comparable with that for the magnetic ions of $d$ or $f$ types \cite{janak}.
Another benefit of the small atomic radius of impurity is its near zero defect formation
energy that shifts the solubility limit to magnitudes as high as $\sim$$10^{22}$ cm$^{-3}$. 
For the spintronics applications, the chemical elements from group 13 of the periodic table 
are found to induce $d_0$ ferromagnetism. 
Among all, in particular the Ga impurity shows the characteristics 
required to establish the magnetic percolation above the room temperature.
Overall, it has been shown that the $\alpha$-PbO crystal structure is a good candidate to 
become the semiconductor host for $d^0$ magnetism as it offers a flexibility to tune 
magnetism by the changing the impurity type.

\section{Acknowledgement}
This work was made possible by the computational facilities of 
Dr. O. Rubel and the Shared Hierarchical Academic Research Computing Network 
(SHARCNET:www.sharcnet.ca) and Compute/Calcul Canada.
Financial support of Ontario Ministry of Research and Innovation through a
Research Excellence Program Ontario network for advanced medical imaging detectors is highly acknowledged.


\begin{thebibliography}{99}
\bibitem{ando}
Ando K., Science {\bf 312} 2006, 1883.

\bibitem{mahadevan}
Mahadevan P. Zunger A. and Sarma D.D. Phys. Rev. Lett. {\bf 93} 2004, 177201.

\bibitem{zunger}
Zunger A., Lany S. and Raebiger H. Physics {\bf 3} 2010, 53.

\bibitem{hu}
Xu H.J. Zhu H.C., Shan X.D., Liu Y.X., Gao J.Y., Zhang X.Z., Zhang J.M., Wang P.W., Hou Y.M. and Yu D.P., 
J. Phys.: Condens. Matter {\bf 22} 2010, 016002.

\bibitem{dietl}
Dietl T., Ohno H., Matsukura F., Cibert J. and Ferrand D.,
Science {\bf 287} 2000, 1019.

\bibitem{gohda}
Y. Gohda Y. and Oshiyama A., 
Phys. Rev. B {\bf 78} 2008, 161201.

\bibitem{coey}
Coey J.M.D.
Solid State Sciences {\bf 7}, 2005, 660.

\bibitem{venk}
Venkatesan M., Fitzgerald C.B. and Coey J.M.D.
Nature {\bf 430} 2004, 630.

\bibitem{podila}
Podila R., Queen W., Nath A., Arantes J. T., Schoenhalz A. L., Fazzio A., Dalpian G. M., He J., Hwu S.J., Skove M. J. and Rao A. M.
Nano Letters {\bf 10} 2010, 1383.

\bibitem{dev}
Dev P., Xue Y. and Zhang P.
Phys. Rev. Lett. {\bf 100} 2008, 117204.

\bibitem{peng}
Peng H., Xiang H. J., Wei S.-H., Li S.-S., Xia J.-B. and Li J. 
Phys. Rev. Lett. {\bf 102} 2009, 017201.

\bibitem{osorio}
Osorio-Guillen J., Lany S., Barabash S.V. and Zunger A.
Phys. Rev. Lett. {\bf 96} 2006, 107203.

\bibitem{khan}
Khan Z. A. and S. Ghosh
Phys. Rev. Lett. {\bf 99} 2011, 042504.

\bibitem{ye}
Ye L.-H.,Freeman A. J. and Delley B.
Phys. Rev. B {\bf 73} 2006, 033203.

\bibitem{shen}
Shen L., Wu R. Q., Pan H., Peng G. W., Yang M., Sha Z.D. and Feng Y.P.
Phys. Rev. B {\bf 78} 2008, 073306.

\bibitem{magnet}
Berashevich J, and Reznik A. arxiv:1304.2945

\bibitem{expmy}
Preliminary results: EPR spectroscopy performed on undoped
PbO samples has revealed the paramagnetic centers in the triplet state at room temperature.

\bibitem{lee}
Lee P.A., Nagaosa N. and Wen X.-G.
Rev. Mod. Phys. {\bf 17} 2006, 78.

\bibitem{mass}
Berashevich J, Semeniuk O., Rowlands J.A. and Reznik A.
EPL {\bf 99} 2012, 47005.

\bibitem{perdew}
Perdew J.P., Burke K. and Ernzernof M.
Phys. Rev. Lett. {\bf 77} 1996, 3865.

\bibitem{wien}
Blaha P., Schwarz K., Madsen G.K.H., Kvasnicka D. and Luitz J.
Wien2k, Techn. Universit\"at Wien, Austria, 2001.

\bibitem{Avezac}
d'Avezac M., Calandra M. and Mauri F.
Phys. Rev. B {\bf 71} 2005, 205210.

\bibitem{exp}
Thangaraju B. and Kaliannann P.
Semicond. Sci. Technol. {\bf 15} 2000, 542.

\bibitem{defects}
Berashevich J., Semeniuk O., Rubel O., Rowlands J.A. and Reznik A.
J. Phys.: Condens. Matter {\bf 25} 2013, 075803.

\bibitem{def}
Berashevich J. and Reznik A. 
J. Phys.:Condens. Matter. {\bf 25} 2013, 475801.

\bibitem{janak}
Janak J. F.
Phys. Rev. B {\bf 16} 1977, 255.

\bibitem{slater}
Slater J. C.
Journal of Chemical Physics {\bf 41} 1964, 3199.

\bibitem{curie}
Ya P., Jin X.F., Kudrnovsky J., Wang D.S. and Bruno P.
Phys. Rev. B {\bf 77} 2008, 054431.

\end{thebibliography}
\end{document}